\documentclass[prl,twocolumn,nopacs,nofootinbib,preprintnumbers]{revtex4}

\usepackage{comment}
\usepackage{mathrsfs}
\usepackage{yfonts}
\usepackage{tipa}
\usepackage{bm}
\usepackage{bbm}
\usepackage{amsmath}
\usepackage{amssymb}
\usepackage{amsthm}
\usepackage{amsfonts}
\usepackage{mathtools}
\usepackage{color}
\usepackage{latexsym}
\usepackage{relsize}
\usepackage[greek,english]{babel}
\usepackage{booktabs}
\usepackage[iso-8859-7]{inputenc}

\usepackage{array}
\newcolumntype{?}{!{\vrule width 2pt}}

\newcommand{\lp}{\left(}
\newcommand{\rp}{\right)}
\newcommand{\lb}{\left[}
\newcommand{\rb}{\right]}

\newcommand{\mQ}{\mathcal{Q}}

\newcommand{\mS}{{\mathcal S}}

\newcommand{\lsim}   {\mathrel{\mathop{\kern 0pt \rlap
  {\raise.2ex\hbox{$<$}}}
  \lower.9ex\hbox{\kern-.190em $\sim$}}}
\newcommand{\gsim}   {\mathrel{\mathop{\kern 0pt \rlap
  {\raise.2ex\hbox{$>$}}}
  \lower.9ex\hbox{\kern-.190em $\sim$}}}
\newcommand{\bw}{\begin{widetext}\begin{equation}}
\newcommand{\ew}{\end{equation}\end{widetext}}
\newcommand{\be}{\begin{equation}}
\newcommand{\ee}{\end{equation}}
\newcommand{\ba}{\begin{eqnarray}}
\newcommand{\ea}{\end{eqnarray}}

\newcommand{\diff}{{{\rm d}}}

\newcommand{\nn}{\nonumber}

\newcommand{\Q}{{\mathcal Q}}

\begin{document}

\title{Coincident General Relativity}

\author{Jose Beltr\'an Jim\'enez$^{a,b}$}\email{jose.beltran@usal.es}
\author{Lavinia Heisenberg$^c$}\email{lavinia.heisenberg@eth-its.ethz.ch}
\author{Tomi Koivisto$^d$}\email{tomik@astro.uio.no}
\address{$^a$Instituto de F\'isica Te\'orica UAM-CSIC, Universidad Aut\'onoma de Madrid, Cantoblanco, Madrid, 28049 Spain}
\address{$^b$Departamento de F\'isica Fundamental, Universidad de Salamanca, E-37008 Salamanca, Spain.}
\address{$^c$Institute for Theoretical Studies, ETH Zurich, Clausiusstrasse 47, 8092 Zurich, Switzerland}
\address{$^d$Nordita, KTH Royal Institute of Technology and Stockholm University, Roslagstullsbacken 23, SE-10691 Stockholm, Sweden}

\preprint{NORDITA-2017-100}
\preprint{IFT-UAM/CSIC-17-093}
\date{\today}

\begin{abstract}

The metric-affine variational principle is applied to generate teleparallel and symmetric
teleparallel theories of gravity. From the latter is discovered an exceptional
class which is consistent with a vanishing affine connection. Based on this  
remarkable property, this work proposes a simpler geometrical formulation of
General Relativity that is oblivious to the affine spacetime structure,
thus fundamentally depriving gravity of any inertial character. The 
resulting theory is described by the Hilbert action purged from the 
boundary term and is more robustly underpinned by the spin-2 field theory,
where an extra symmetry is now manifest, possibly related to the double copy structure of the gravity amplitudes. This construction also provides a novel starting point for modified gravity 
theories, and the paper presents new and simple generalisations where 
analytical self-accelerating cosmological solutions arise naturally in the 
early and late time universe. 

\end{abstract}

\maketitle

In the conventional geometrical interpretation of General Relativity (GR), gravitation is described by the curvature 
\be \label{riemann}
{R}^\alpha_{\phantom{\alpha}\beta\mu\nu} \equiv 
2\partial_{[\mu} \Gamma^\alpha_{\phantom{\alpha}\nu]\beta}
+ 2\Gamma^\alpha_{\phantom{\alpha}[\mu\lvert\lambda\rvert}\Gamma^\lambda_{\phantom{\lambda}\nu]\beta}\,,
\ee
where the affine connection $\Gamma^\alpha{}_{\mu\nu}=\{^{\phantom{i} \alpha}_{\beta\gamma}\}$ is the metric-compatible and torsionless Levi-Civita connection that is computed from the 
metric $g_{\mu\nu}$ (with mostly plus Lorentizan signature) as
\be \label{christoffel}
\left\{^{\phantom{i} \alpha}_{\beta\gamma}\right\} \equiv \frac{1}{2}g^{\alpha\lambda}\lp g_{\beta\lambda,\gamma}
+  g_{\lambda\gamma,\beta} - g_{\beta\gamma,\lambda}\rp\,.
\ee
Despite its undeniable observational success, this description comes hand in hand with the inherent conceptual and technical difficulties of working in a pseudo-Riemannian spacetime so that it is desirable to attain a simpler formulation of the theory.
A step forward in this direction is taken in the teleparallel reformulation \cite{Aldrovandi:2013wha,Maluf:2013gaa}, where the geometry is simplified by the constraint ${R}^\alpha_{\phantom{\alpha}\beta\mu\nu} = 0$, which reduces the connection to the Weizenb\"ock form. In this work we will pursue further improvements in the simplifying sequence of geometrical frameworks, and formulate GR in a flat and torsion-free spacetime. Furthermore, we will introduce a class of its generalisations in the trivially connected geometry, i.e with $\Gamma^\alpha_{\phantom{\alpha}\mu\nu}=0$. The standard vierbein formalism of GR has been reinterpreted in terms of nonlinear realisations of the group $GL(4,\mathbb{R})$ \cite{1971AnPhy..62...98I}, and teleparallelised in the metric-affine gauge theory (\cite{Obukhov:2002tm,Nester:2017wau}, 5.9 in \cite{Hehl:1994ue}), but it can still be clarified whether it is natural to stipulate a trivial geometry since there, as we shall discover, the inertial connection is a 
translation\footnote{Compelling arguments to regard gravitation as the gauge theory of translations were presented e.g. in Ref. \cite{Feynman:1996kb}, wherein also the field theoretical 
approach as an alternative to the conventional geometrical approach to gravitation was emphasised. Note that in the Weitzenb\"ock teleparallelism \cite{Aldrovandi:2013wha,Maluf:2013gaa}, gravitation is still geometry, and despite the motivation as a translation gauge theory, the gauge connection is a (pseudo-)rotation.}. 

We start by recalling that the general affine connection admits the decomposition \cite{Ortin:2015hya} 
\be \label{decomposition}
\Gamma^\alpha_{\phantom{\alpha}\mu\nu}=\left\{^{\phantom{i} \alpha}_{\mu\nu}\right\} + K^\alpha_{\phantom{\alpha}\mu\nu} + L^\alpha_{\phantom{\alpha}\mu\nu}\,,
\ee
which includes the {\it contortion}
\be 
K^\alpha_{\phantom{\alpha}\mu\nu} \equiv \frac{1}{2}T^\alpha_{\phantom{\alpha}\mu\nu} + T_{(\mu{\phantom{\alpha}\nu)}}^{\phantom{,\mu}\alpha}\,,
\ee
due to torsion $T^\alpha_{\phantom{\alpha}\mu\nu} \equiv 2\Gamma^\alpha_{\phantom{\alpha}[\mu\nu]}$, and the {\it disformation}
\be \label{disformation}
L^\alpha_{\phantom{\alpha}\mu\nu} \equiv  \frac{1}{2} Q^{\alpha}_{\phantom{\alpha}\mu\nu} - Q_{(\mu\phantom{\alpha}\nu)}^{\phantom{(\mu}\alpha}\,,
\ee
due to non-metricity $Q_{\alpha\mu\nu} \equiv \nabla_\alpha g_{\mu\nu}$. 
The various tensor fields we have already introduced satisfy important identities, an example being the Bianchi identity \cite{Ortin:2015hya}
\be
R^\mu_{\phantom{\mu}[\alpha\beta\gamma]} - \nabla_{[\alpha}T^\mu_{\phantom{\mu}\beta\gamma]} + T^\nu_{\phantom{\nu}[\alpha\beta}T^\mu_{\phantom{\mu}\gamma]\nu} =  0\,, \label{bianchi2}
\ee
which we shall need later. 
In the metric-affine formalism, a theory can be now defined by a scalar action of the form
\be \label{action}
\mS = \int \diff^n x \sqrt{-g}f(g^{\mu\nu},R^\alpha_{\phantom{\alpha}\beta\mu\nu},T^\alpha_{\phantom{\alpha}\mu\nu},Q_{\alpha}^{\phantom{\alpha}\mu\nu}) + \mathcal{S}_M\,,
\ee
where the independent variables are the metric and the affine connection, plus the matter fields contained in $\mS_M$. The sources introduced by the latter are the energy-momentum tensor and the hypermomentum tensor density defined as
\be 
\mathfrak{T}_{\mu\nu} = -\frac{2}{\sqrt{-g}}\frac{\delta \mathcal{S}_M }{\delta g^{\mu\nu}}\,, \quad 
\mathfrak{H}_\lambda{}^{\mu\nu} = -\frac12\frac{\delta \mathcal{S}_M}{\delta \Gamma^\lambda_{\phantom{\lambda}\mu\nu}}\,,
\ee
respectively. 

\section{The new GR}

Before proceeding to the central result of this work, it will be interesting to rederive the so-called New GR \cite{Hayashi:1979qx} covariantly in the metric-affine formalism. If we introduce the {\it superpotential}   
\be
{S}{}_\alpha^{\phantom{\alpha}\mu\nu}   \equiv  a T_\alpha^{\phantom{\alpha}\mu\nu} + b T^{[\mu\phantom{\alpha}\nu]}_{\phantom{,\mu}\alpha} + c\delta^{[\mu}_\alpha T^{\nu]}\,, \label{super}
\ee
then the general even-parity quadratic theory can be defined in terms of the three-parameter quadratic form $\mathbb{T} \equiv \frac{1}{2}{S}{}_\alpha^{\phantom{\alpha}\mu\nu}T^\alpha_{\phantom{\alpha}\mu\nu}$. Thus, instead of imposing a priori the flatness and metricity conditions on the connection, we set up the gravitational action as   
\be \label{abcd}
\mS = \int \diff^n x \lb  \frac12\sqrt{-g}\mathbb{T}  
+ 
\lambda_\alpha^{\phantom{\alpha}\beta\mu\nu} R^\alpha_{\phantom{\alpha}\beta\mu\nu} + \lambda^\alpha_{\phantom{\alpha}\mu\nu}\nabla_\alpha g^{\mu\nu} \rb\,,
\ee
where the connection is left completely arbitrary and the Lagrange multipliers that impose the teleparallelism constraints are introduced as tensor densities with the obvious symmetries
$\lambda_\alpha{}^{\mu\nu}=\lambda_\alpha{}^{(\mu\nu)}$ and $\lambda_\alpha{}^{\nu\mu\rho}=\lambda_\alpha{}^{\nu[\mu\rho]}$. We first compute the field equations by varying the action (\ref{abcd}) w.r.t. the metric, yielding
\ba
&&T_{\alpha\beta(\mu}S^{\alpha\beta}_{\phantom{\alpha\beta}\nu)}-\frac12T_{(\mu\lvert\alpha\beta\rvert}S_{\nu)}^{\phantom{\nu}\alpha\beta} -\frac12g_{\mu\nu}\mathbb{T} \nn\\ &&\quad=\mathfrak{T}_{\mu\nu} +  \frac{2}{\sqrt{-g}}\lp \nabla_\alpha + T_{\alpha}\rp \lambda^{\alpha}_{\phantom{\alpha}\mu\nu}\,, \label{efes}
\ea
while the variations w.r.t. the connection result in 
\ba \label{eomlambda}
\Big(\nabla_\rho+T_\rho\Big) \lambda_{\alpha}{}^{\nu\mu\rho}+\frac12 T^\mu{}_{\rho\sigma}\lambda_\alpha{}^{\nu\rho\sigma}=\Delta_\alpha{}^{\mu\nu}\,,
\ea
where we have defined the source term
\be \label{delta}
\Delta_\alpha{}^{\mu\nu}\equiv \mathfrak{H}_\alpha{}^{\mu\nu}+\frac{1}{2}\sqrt{-g}S_\alpha{}^{\nu\mu}-\lambda^{\mu\nu}{}_\alpha\,,
\ee
and used the symmetries of $\lambda_{\alpha}{}^{\nu\mu\rho}$ and $\lambda_\alpha{}^{\mu\nu}$. Now the field equations (\ref{efes}) seem to have no dynamics
for the torsion. However, we still need to solve for the constraints imposed by the Lagrange multipliers. Using the constraint ${R}^\alpha_{\phantom{\alpha}\beta\mu\nu} = 0$ and the antisymmetry
of the corresponding Lagrange multiplier, we can derive the divergence of the source term as
\be \label{eomlambdaJ2}
\frac12\nabla_\mu T^\mu{}_{\rho\sigma}\lambda_\alpha{}^{\nu\rho\sigma}-\nabla_\mu T_\rho\lambda_\alpha{}^{\nu\rho\mu}-T_\rho\nabla_\mu\lambda_\alpha{}^{\nu\rho\mu}=\nabla_\mu\Delta_\alpha{}^{\mu\nu}\,. 
\ee
Then we can use (\ref{eomlambda}) to replace $\nabla_\mu\lambda_\alpha{}^{\nu\rho\mu}$, and yet take advantage of the Bianchi identity (\ref{bianchi2}) to
verify that 
\be 
\Big(\nabla_\mu+T_\mu\Big)\Delta_\alpha{}^{\mu\nu}=0\,. \label{eqDivlambda}
\ee
We can then plug this result in (\ref{efes}) to rewrite them as
\ba
 \mathfrak{T}_{\mu\nu} & +&  \frac{2}{\sqrt{-g}}(\nabla_\alpha + T_\alpha)\mathfrak{H}^{\phantom{\mu}\alpha}_{\mu\phantom{\alpha}\nu} =  -\mathcal{D}_\alpha {S}_{\mu\nu}{}^\alpha \nn \\
& - &   S^{\alpha\beta}{}_\nu\left(T_{\alpha\mu\beta}+K_{\alpha\beta\mu} \right)
 -  \frac12g_{\mu\nu}\mathbb{T}\,, \label{efes3}
\ea
where we have referred to the familiar Levi-Civita covariant derivative $\mathcal{D}_\alpha$ that comes with the Christoffel symbols (\ref{christoffel}). 
Note that we have removed the symmetrisation from (\ref{efes3}) as a more compact manner of writing all the equations: the 10 symmetric components correspond to the metric field equations and, in general, the equation has also 6  antisymmetric components which are nothing but the equations (\ref{eqDivlambda}). In the standard prescription
for standard matter fields, the hypermomentum decouples from the symmetric equations. The teleparallel equivalent of GR (TEGR) is reproduced by the choice of parameters 
$a=\frac{1}{4}$, $b=\frac{1}{2}$, $c=-1$ and $n=4$. 

We have now seen that this theory and its generalisations can be formulated without introducing the additional structure of the frame bundle and its corresponding extra set of indices. The connection can then be reduced to its contortion component entirely determined by the torsion, which is propagating due to the restriction to a teleparallel geometry\footnote{Due to the teleparallelity constraint, the connection includes the first derivative of a general linear transformation, and the second derivatives resulted in the field equation (\ref{efes3}) from solving multiplier that imposes the metricity constraint. The gauge freedom of the
Lagrange multipliers and the number of effective components of the
multipliers and equations have been investigated previously in the gauge formalism \cite{Blagojevic:2002du,Obukhov:2002tm,Nester:2017wau}.}.

\section{A newer GR}

We then move to the much less explored case of symmetric teleparallelisms \cite{Nester:1998mp}, see also \cite{Adak:2005cd,Adak:2006rx,Adak:2008gd,Mol:2014ooa}. The physical interpretation we suggest extrapolates the successful argument of teleparallelism to its logical conclusion. As it is well-known, GR cannot distinguish between gravitation and inertial effects, 
but by resorting to frame fields, the gravitational energy can be defined covariantly in the teleparallel approach \cite{Moller,deAndrade:2000kr,Maluf:2013gaa}. 
The canonical frame is now identified by the absence of curvature and torsion, and the canonical coordinates are now identified by the absence of inertial effects. This is a physical rationale how to 
extract quantities of interest such as the gravitational energy and the gravitational entropy, and how to proceed with the quantisation and with the unification. The aim is to establish the frame and the coordinate system wherein the canonical commutation relations can be recovered for the operators corresponding to physical observables.

The non-metricity tensor has two independent traces, which we denote as $Q_\mu=Q_{\mu\phantom{\alpha}\alpha}^{\phantom{\mu}\alpha}$ and
$\tilde{Q}^\mu = Q_\alpha^{\phantom{\alpha}\mu\alpha}$. We can then define the quadratic non-metricity scalar as
\be  \label{qdef}
\Q =   -\frac{1}{4}Q_{\alpha \beta \mu}Q^{\alpha\beta \mu} +  \frac{1}{2}Q_{\alpha \beta \mu}Q^{\beta \mu\alpha} 
  +   \frac{1}{4}Q_\alpha Q^\alpha  
  - \frac{1}{2}Q_\alpha\tilde{Q}^\alpha\,
\ee
that is special among the general quadratic combination because, in addition to being invariant under local general linear transformations, it is also the special quadratic combination that is invariant under a translational symmetry that allows to completely remove the connection. 
To clarify this remarkable property, let us introduce, in analogy with the superpotential of New GR, the following {\it non-metricity conjugate}
\ba
P^\alpha{}_{\mu\nu} & \equiv & c_1 Q^\alpha{}_{\mu\nu}+c_2Q_{(\mu\phantom{\alpha}\nu)}^{\phantom{\mu}\alpha}+c_3 Q^\alpha g_{\mu\nu} \nonumber \\
& + & c_4\delta^\alpha_{(\mu}\tilde{Q}_{\nu)} + \frac{c_5}{2}\big(\tilde{Q}^\alpha g_{\mu\nu}+\delta^\alpha_{(\mu}Q_{\nu)}\big)\,,
\ea
and define the general quadratic form $\mathbb{Q} = Q_\alpha{}^{\mu\nu}P^\alpha{}_{\mu\nu}$. Then,
one may consider the general quadratic theory
\be  \label{abcd4}
\mS  =   \int \diff^n x\Big[-\frac{1}{2}\sqrt{-g}\mathbb{Q}
 +  \lambda_\alpha^{\phantom{\alpha}\beta\mu\nu} R^\alpha_{\phantom{\alpha}\beta\mu\nu} + \lambda_\alpha^{\phantom{\alpha}\mu\nu}T^\alpha_{\phantom{\alpha}\mu\nu}\Big]\,.
\ee
Noteworthy, the five terms that go into the definition of the full conjugate can be related to the squares of the four irreducible pieces of the non-metricity and their one possible cross-term \cite{Hehl:1994ue}. The special non-metricity scalar given in (\ref{qdef}) corresponds to the choice of parameters $c_4=0$ and $c_1 = -\frac{1}{2}c_2=-c_3= \frac{1}{2}c_5=-\frac{1}{4}$, i.e., in that case we have $\mathbb{Q}=\Q$ and the theory simply becomes the symmetric teleparallel equivalent of GR (STEGR) \cite{Nester:1998mp,Adak:2005cd,Adak:2006rx,Adak:2008gd,Mol:2014ooa}, as we will explicitly show below. Before that, let us give the field equations for the general case. Variations w.r.t. the metric lead to the equations
\be \label{qefe}
\frac{2}{\sqrt{-g}}\nabla_\alpha (\sqrt{-g}P^\alpha{}_{\mu\nu}) - q_{\mu\nu} -  \mathbb{Q} g_{\mu\nu}= \mathfrak{T}_{\mu\nu}\,,
\ee
where the short-hand $q_{\mu\nu}$ stands for 
\ba q_{\mu\nu}
&=&c_1\Big(2Q_{\alpha\beta\mu}Q^{\alpha\beta}{}_\nu-Q_{\mu\alpha\beta}Q_\nu{}^{\alpha\beta}\Big)\nonumber\\
&+&c_2Q_{\alpha\beta\mu}Q^{\beta\alpha}{}_\nu+c_3\Big(2Q_\alpha Q^\alpha{}_{\mu\nu}-Q_\mu Q_\nu\Big)\nonumber\\
&+&c_4\tilde{Q}_\mu \tilde{Q}_\nu+c_5\tilde{Q}_\alpha Q^\alpha{}_{\mu\nu}\,.
\ea
On the other hand, the connection equations are
\be \label{cefe}
\nabla_\rho\lambda_\alpha{}^{\nu\mu\rho}+\lambda_\alpha{}^{\mu\nu}= \sqrt{-g}P^{\mu\nu}{}_\alpha+\mathfrak{H}_\alpha{}^{\mu\nu}\,.
\ee
Notice that the Lagrange multipliers do not enter the metric field equations (\ref{qefe}), which is a technical simplification as compared to the teleparallel geometry with torsion considered in the previous section. Thus, the gauge symmetries of the multipliers \cite{Blagojevic:2002du,Obukhov:2002tm,Nester:2017wau} in this case are irrelevant (see \cite{BeltranJimenez:2018vdo}). The vanishing curvature constraint imposes the connection to be purely inertial, i.e., it differs from the trivial connection by a general linear gauge transformation, while the torsionless condition further simplifies it to take the form $\Gamma^\alpha{}_{\mu\beta}=(\partial x^\alpha/\partial \xi^\lambda)\partial_\mu\partial_\beta\xi^\lambda$ for some arbitrary $\xi^\lambda$. We then arrive at the crucial result that we can completely remove the connection by means of a diffeomorphism (called a ''Diff'' hereafter) and, thus, the $\xi^\lambda$'s make their appearance as the St\"uckelberg fields restoring this gauge symmetry. If we fix the gauge $\Gamma^\alpha_{\phantom{\alpha}\mu\nu}=0$ and expand the Lagrangian to quadratic order in the perturbations $g_{\mu\nu} = \eta_{\mu\nu}+h_{\mu\nu}$, we obtain
\ba
\sqrt{-g}\mathcal{L}^{(2)} &=&c_1\partial_\alpha h_{\mu\nu} \partial^\alpha h^{\mu\nu}+(c_2+c_4)\partial_\alpha h_{\mu\nu}\partial^\mu h^{\alpha\nu}
\nonumber\\
&&+c_3\partial_\alpha h \partial^\alpha h+c_5\partial_\mu h^\mu{}_\nu\partial^\nu h\,.
\label{eq:h2}
\ea
As it is well-known, this Lagrangian propagates more than two dof's unless a linearised Diff-symmetry is imposed\footnote{Actually, there is a second option where only transverse Diffs are imposed together with an additional Weyl rescaling.}. The requirement of this Diff-symmetry thus fixes the theory, up to a degeneracy of $c_2+c_4$ and the overall normalisation, to the case of the symmetric teleparallel equivalent of GR, i.e. $\mathbb{Q}=\Q$. For that choice of parameters, the trivial connection is consistent with maintaining Diff-invariance because the inertial connection drops from the non-metricity sector of the action. An extra Diff was found to be at work, in the self-dual sector \cite{Monteiro:2011pc}, making possible the double-copy structure of the gravity amplitudes \cite{Bern:2010ue}.
Let us also notice that (\ref{eq:h2}) also gives directly the theoretical (absence of instabilities) and phenomenological (Newtonian limit compatible with Solar System observations) constraints of the general quadratic theory.

\section{The f(Q) cosmology}

The new simple geometrical formulation of GR motivates a promising new framework for studies of modified gravity. To demonstrate this with an example, we propose the following action: 
 \be  \label{abcd3}
\mS_G  =   \int \diff^n x\Big[\frac{1}{2}\sqrt{-g} f(\Q)
 +  \lambda_\alpha^{\phantom{\alpha}\beta\mu\nu} R^\alpha_{\phantom{\alpha}\beta\mu\nu} + \lambda_\alpha^{\phantom{\alpha}\mu\nu}T^\alpha_{\phantom{\alpha}\mu\nu}\Big]\,.
\ee
where the special case $f=\Q$ corresponds to the Einstein action as discussed above. For cosmological applications we consider the line element $\diff s^2 = -\diff t^2 + a^2(t)\delta_{ij}\diff x^i \diff x^j$ with the expansion rate defined as $H=\dot{a}/a$. In the coincident gauge with $\Gamma^\alpha_{\phantom{\alpha}\mu\nu}=0$ we have $\mQ=6H^2$ and the cosmological equations in the presence of a perfect fluid with energy density $\rho$ and pressure $p$ are, with the Newton's constant $G$ restored, 
\ba
6f'H^2-\frac{1}{2}f& = & 8\pi G\rho\,, \label{fried1} \\
\lp 12f''H^2 + f'\rp\dot{H} & = & -4\pi G\lp\rho+p\rp\, . \label{fried2}
\ea
It is straightforward to verify that the continuity equation $\dot{\rho}=-3H(\rho+p)$ is sustained by the Bianchi identities applied to the LHS of the above equations, which is a cross-check of the consistency of our gauge choice and Ansatz for the cosmological solution\footnote{Notice that we have used the Diff gauge freedom to fix the coincident gauge and, therefore, setting the lapse to 1 is not, in principle, a permitted choice. It happens however that the $f(\mQ)$ theories retain a time-reparameterisation invariance that allows to get rid of the lapse (see \cite{BeltranJimenez:2018vdo}).}. 

One immediate interesting observation is that the cosmological background evolution of the theory $f=\mQ+\Lambda\sqrt{\mQ}$, where $\Lambda$ is some scale, cannot be distinguished from GR,
but the perturbation evolution could be studied to constrain the value of $\Lambda$.

In order to grasp some interesting phenomenological consequences of general $f(\mQ)$ theories, let us consider $f(\mQ)=\mQ-6\lambda M^2(\frac16\mQ/M^2)^\alpha$ with $M$ some scale and $\lambda$ a dimensionless parameter (that we will assume to be order 1).  The modified Friedmann equation reads 
 \be
 H^2\left[1+(1-2\alpha)\lambda\left(\frac{H^2}{M^2}\right)^{\alpha-1}\right]=\frac{8\pi G}{3}\rho.
 \ee
For $\alpha=1/2$ we again recover the usual GR background cosmology. We then see that for $\alpha>1$ the GR evolution is recovered for $H^2\ll M^2$ and the modifications appear in the early universe whenever $H^2\gtrsim M^2$. The evolution in that regime becomes $H^2\propto\rho^{\frac{1}{\alpha}}$ which shows a potential inflationary solution for $\alpha$ sufficiently large. Notice that this transition is only possible if $(1-2\alpha)\lambda>0$. Interestingly, if this is not the case the evolution can lead to a maximum value for $H^2$ and $\rho$. In general we expect to have several branches of solutions and it seems a reasonable condition to choose the one matching GR at low densities/curvatures. On the other hand, if $\alpha<1$ we expect the modifications to become relevant in the late universe when $H^2\lsim M^2$ so that these models can provide self-accelerating solutions where the universe transits from a matter dominated epoch to an asymptotically de Sitter universe.

These promising cosmological scenarios deserve a more detailed analysis, which we leave, together with the perturbation analysis of general $f(\Q)$ models, to a future work\footnote{See also \cite{Jimenez:2015fva} and \cite{Jimenez:2016opp} for related cosmological solutions in the framework of vector distortion, where the connection (\ref{christoffel}) was amended with three vector-field-dependent terms with the intention to parameterise the effects of generalised gauge geometry.}. 

\section{Conclusion: The purified GR}

We hope that the presented formulation is useful in linking some concepts of generalised spacetime geometry with experimental precision science. We demonstrated such a possibility in the field of cosmology, by sketching how a strikingly simple class of models that exists in the symmetric teleparallel geometry \cite{Nester:1998mp,Adak:2008gd,Mol:2014ooa} can provide potentially viable alternatives to inflation and dark energy. In the following we show further advantages of the purified framework in some specific applications:

\paragraph{Gauge theory.} 
In the gauge with vanishing connection, the non-metricity scalar $\Q$ can be expressed in terms of the Christoffel symbols (\ref{christoffel}) as
\be \label{gg}
\Q=g^{\mu\nu}\Big(\left\{^{\phantom{i} \alpha}_{\beta\mu}\right\} \left\{^{\phantom{i} \beta}_{\nu\alpha}\right\} -\left\{^{\phantom{i} \alpha}_{\beta\alpha}\right\}\left\{^{\phantom{i} \beta}_{\mu\nu}\right\} \Big). 
\ee
The RHS of this relation is nothing but the quadratic piece of the Hilbert action\footnote{Which is often known as the ''Einstein-Hilbert action''. However, some authors \cite{Chen:2018vce} refer to the scalar curvature action
as the Hilbert action, and we can then say that Einstein action, i.e. the ''$\Gamma\Gamma$-action'' given by (\ref{gg}), differs from the Hilbert action by a total derivative. A covariant version of the  ''$\Gamma\Gamma$-action'' has also been constructed by the means of a reference connection associated to a reference metric \cite{Tomboulis:2017fim}.}
 that only differs from the Ricci scalar by a total derivative so that, written in terms of the metric, both actions are equivalent. Thus, we arrive at the remarkable result that the theory described by $\Q$ is equivalent to an improved version of GR where the boundary term is absent, 
the connection can be fully trivialised and, thus, the inertial character of gravity as an effect of the connection has completely disappeared, which represents a much simpler geometrical interpretation of gravity. Furthermore, the trivial connection corresponds to the unitary gauge for the St\"uckelbergs $\xi^\lambda=x^\lambda$ so that the origins of the tangent space and the spacetime coincide.Therefore we call this theory the {\it coincident GR.} The covariant $\Gamma\Gamma$ action realises gravitation as a gauge theory of translations.

\paragraph{Field theory.} Another notable feature of this formulation of GR is that it exactly reproduces the resummation for a self-interacting massless spin 2 field \cite{Deser:1969wk} (see also \cite{Padmanabhan:2004xk,Butcher:2009ta,Barcelo:2014mua,Tomboulis:2017fim}), unlike GR where the boundary term must be added by hand,
as was pointed out in Ref.\cite{Padmanabhan:2004xk}. The coincident GR thus provides a more robust relation with the field theory approach to gravity. In addition, the existence of a gravitational energy-momentum tensor \cite{Moller,deAndrade:2000kr,Maluf:2013gaa,Nester:1998mp,Chen:2018vce} makes this approach less ambiguous (this does not get around the Weinberg-Witten obstruction though \cite{Weinberg:1980kq}). Finally, we may remind that in the loop computations, clear technical advantages of the $\Gamma\Gamma$ action with respect to the Hilbert action have already been known, and as pointed out above, our covariant realisation further hints at a understanding of the double copy structure \cite{Bern:2010ue,Monteiro:2011pc}.

\paragraph{Euclidean action.} The coincident GR Lagrangian may also provide new insights into the Euclidean approach to quantum gravity. For instance, in the usual GR computation of the black hole entropy, the full contribution is given by the Gibbons-Hawking-York boundary term which, in addition, needs to be properly normalised to obtain a finite result. In contrast, the coincident GR Lagrangian requires neither a boundary term nor regularisation, but instead allows to identify a set of natural coordinates\footnote{Notice that Diffs are only realised up to a boundary term in the action of coincident GR and, therefore, the value of the Euclidean action will depend on the choice of coordinates (or rather on classes of coordinates non-trivially related at infinity).} which directly give a finite value (see \cite{BeltranJimenez:2018vdo} for more details). 

\paragraph{Matter couplings.} In non-Riemannian geometries, ambiguities arise regarding the coupling of matter. In spacetimes with torsion, the gravitational coupling of gauge fields such as 
the Maxwell field can spoil the gauge invariance of the theory. This is obviously avoided by our torsion-free $\nabla$.
In TEGR, a further problem occurs with fermions, since they couple to the axial contorsion of the Weitzenb\"ock connection. 
This is not viable, and one has to ad hoc invoke the usual Levi-Civita connection and proclaim that
it defines the matter coupling also in TEGR. In the STEGR however, this problem is completely avoided due to the property that Dirac fermions only couple to the completely antisymmetric part of the affine connection and, thus, they are oblivious to any disformation piece. The viability of minimal coupling gives further support to our claim that the coincident GR represents the translation gauge theory
in the unique geometry purified from inertial effects.

Conceptually, TEGR \cite{Aldrovandi:2013wha,Maluf:2013gaa} has offered a tensor for the gravitational energy-momentum \cite{Moller,deAndrade:2000kr,Maluf:2013gaa,Nester:1998mp}, and currently new insights are sought into holography and entropy, see e.g. \cite{Krssak:2015rqa}. Meanwhile modified non-equivalent theories are vigorously investigated in the context of cosmology \cite{Cai:2015emx}.

Having arrived at a simpler and possibly yet more consistent formulation of GR, we may conclude this work with an open mind for a new convention in geometry.

\vspace{0.3cm}
{\bf Acknowledgments}: JBJ acknowledges the financial support
of A*MIDEX project (n ANR-11-IDEX-0001-02) funded by the Investissements d'Avenir French Government program, managed by the French National Research Agency (ANR), MINECO (Spain) projects FIS2014-52837-P, FIS2016-78859-P (AEI/FEDER), Consolider-Ingenio MULTIDARK CSD2009-00064, Centro de Excelencia Severo Ochoa Program SEV-2016-0597 and the Programa de atracci\'on del talento cient\'ifico funded by the FSCCSS. L.H. acknowledges financial support from Dr. Max R\"ossler, the Walter Haefner Foundation and the ETH Zurich Foundation. This article is based upon work from COST Action CA15117.

\bibliography{fQ}

\end{document}